\documentclass[aps,prb,twocolumn,amsfonts]{revtex4-1}

\usepackage{graphicx}
\usepackage{amssymb}
\usepackage{amsmath}
\usepackage{relsize}
\usepackage{multirow}
\usepackage{bigdelim}
\usepackage{dsfont}

\usepackage[colorlinks=true,citecolor=blue,linkcolor=magenta]{hyperref}
\hypersetup{
pdfauthor = {A. Barfuss},
colorlinks = true, linkcolor = blue, urlcolor=blue, bookmarksnumbered =  true}

\newcommand{\ket}[1]{\left|#1\right>}

\begin{document}


\title{Spin-stress and spin-strain coupling in diamond-based hybrid spin oscillator systems}


\author{A. Barfuss}
\affiliation{Department of Physics, University of Basel, Klingelbergstrasse 82, CH-4056 Basel, Switzerland}
\author{M. Kasperczyk}
\affiliation{Department of Physics, University of Basel, Klingelbergstrasse 82, CH-4056 Basel, Switzerland}
\author{J. K\"olbl}
\affiliation{Department of Physics, University of Basel, Klingelbergstrasse 82, CH-4056 Basel, Switzerland}
\author{P. Maletinsky}
\affiliation{Department of Physics, University of Basel, Klingelbergstrasse 82, CH-4056 Basel, Switzerland}
\email{patrick.maletinsky@unibas.ch}


\date{\today}

\begin{abstract}
Hybrid quantum systems, which combine quantum-mechanical systems with macroscopic mechanical oscillators, have attracted increasing interest as they are well suited as high-performance sensors or transducers in quantum computers. 
A promising candidate is based on diamond cantilevers, whose motion is coupled to embedded Nitrogen-Vacancy (NV) centers through crystal deformation. 
Even though this type of coupling has been investigated intensively in the past, several inconsistencies exist in available literature, and no complete and consistent theoretical description has been given thus far.
To clarify and resolve these issues, we here develop a complete and consistent formalism to describe the coupling between the NV spin degree of freedom and crystal deformation in terms of stress, defined in the crystal coordinate system $XYZ$, and strain, defined in the four individual NV reference frames. 
We find that the stress-based approach is straightforward, yields compact expressions for stress-induced level shifts and therefore constitutes the preferred approach to be used in future advances in the field. 
In contrast, the strain-based formalism is much more complicated and requires extra care when transforming into the employed NV reference frames.
Furthermore, we illustrate how the developed formalism can be employed to extract values for the spin-stress and spin-strain coupling constants from data published by Teissier \textit{et al.} \cite{Teissier2014}.
\end{abstract}

\maketitle

\section{Introduction}
\label{sec:Introduction}
Hybrid systems combine quantum mechanical two-level systems with macroscopic mechanical oscillators and have attracted increasing attention recently, largely with the goal of employing them as high-performance nanoscale sensors or transducers in multi-qubit networks\cite{Treutlein2014}.
Such systems can furthermore serve as testbeds to study macroscopic objects in the quantum regime, provided the coupling between resonator motion and the two-level system resides in the high cooperativity regime \cite{Lee2017}.
An extensive variety of hybrid systems are already being studied. These include superconducting circuits coupled capacitively \cite{LaHaye2009, OConnell2010, Pirkkalainen2013}, ultracold atoms linked by radiation pressure forces \cite{Hunger2010, Camerer2011, Jockel2015} and quantum dots or solid-state spins coupled by magnetic field gradients \cite{Rugar2004, Arcizet2011, Kolkowitz2012} or crystal stress \cite{Yeo2014, Montinaro2014, Munsch2016, MacQuarrie2013, Teissier2014, Ovartchaiyapong2014} to mechanical oscillators of different materials and shapes. 

Hybrid spin-oscillator systems, in which the motion of a diamond resonator is coupled to the spin-degree of freedom of an embedded Nitrogen-Vacancy (NV) color center, are of particular interest \cite{Lee2017}.
The NV center provides a promising solid-state platform for quantum technologies due to its room-temperature operation with long spin coherence times \cite{Balasubramanian2009} and well established optical methods for spin initialization and readout \cite{Gruber1997}.
Diamond resonators benefit from the material's high Young's modulus, which provides excellent mechanical strength and gives rise to exceptional stress amplitudes per resonator displacement. Recent advances in diamond fabrication have further demonstrated high-quality resonators with quality factors $Q\sim10^6$ \cite{Ovartchaiyapong2012, Tao2013, Tao2014}.
Additionally, coupling between resonator and NV spin is intrinsic. Diamond-based hybrid systems thus come with minimized fabrication complexity and immediately offer a robust qubit-resonator link, which is crucial for operation such systems in the quantum regime \cite{Rabl2010, Habraken2012, Kepesidis2013}.

Consequently, the coupling between NV spin and crystal deformation has been explored in various experiments, starting with seminal work by Davies and Hamer, who investigated its influence on the NV's optical transitions in the 1970s \cite{Davies1976}. 
Subsequent studies first aimed at probing the electronic level structure of NV ground and excited states \cite{Tamarat2008, Batalov2009, Teissier2014, Ovartchaiyapong2014, Rogers2008, Doherty2014, Rogers2015}, and recently started to investigate diamond-based hybrid spin-oscillator systems. Substantial evidence was found that quantum ground state operation is in principle possible \cite{Wilson-Rae2004, Rabl2009, Rabl2010a, MacQuarrie2013, Teissier2014, Ovartchaiyapong2014, Meesala2016, MacQuarrie2017}. 
It was also discovered that crystal deformation allows for coherent control of the NV's spin degree of freedom \cite{Barfuss2015, MacQuarrie2015} and that such hybrid systems can have future sensing applications, for example in protecting NV centers from environmental noise through dynamical decoupling \cite{Barfuss2015, MacQuarrie2015a, Teissier2017} or as the main ingredient of spin-mechanical sensors for mass spectrometry and force microscopy \cite{Barson2017}.

Yet even though the coupling between NV spin and crystal deformation has been studied intensively, several differences and inconsistencies in its formal description exist in the published literature.
Crystal deformation is treated in terms of stress \cite{Grazioso2013, MacQuarrie2013, Rogers2015, Barson2017} or strain \cite{Teissier2014, Ovartchaiyapong2014, Meesala2016}, defined in crystal\cite{Grazioso2013, Rogers2015, Barson2017} or defect coordinate systems \cite{MacQuarrie2013, Teissier2014, Ovartchaiyapong2014, Lee2016, Meesala2016}. This already confusing situation is further complicated by inconsistent sign conventions for stress and strain \cite{Teissier2014, Ovartchaiyapong2014, Doherty2014, Barson2017}, and the use of different, occasionally incorrect, interaction Hamiltonians in literature. While recent works employ correct approximations of the complete interaction Hamiltonian \cite{Rogers2015, Barson2017}, earlier studies rely on oversimplified versions where the tensorial nature of strain is neglected \cite{MacQuarrie2013, Teissier2014, Ovartchaiyapong2014}.

To clarify and resolve existing inconsistencies, in this paper we provide a complete and consistent theoretical treatment of the coupling between crystal deformation and the spin degree of freedom of negatively charged NV centers. 
We start the first part with a clear definition of the crystal and NV coordinate systems we employ. Subsequently, we use the recently formulated, complete spin-stress interaction Hamiltonian \cite{Udvarhelyi2017} and calculate stress-induced spin sublevel shifts with the stress tensor defined in crystal coordinates. Here we focus in particular on the question of how to best include all four possible NV orientations in the calculations. We then convert stress- into strain-induced level shifts, with the strain tensors defined in NV coordinate systems, and calculate the corresponding level shifts of the NV spin sublevels. 
In the second part, we derive the stress tensor in a singly-clamped diamond cantilever under the influence of an external shear force. We then use this stress tensor and the developed coupling formalism to illustrate how spin-stress coupling is quantified experimentally. In particular, we reanalyze experimental data from Teissier \textit{et al.} \cite{Teissier2014} and correctly quantify the spin-stress coupling constants.

\section{Spin-stress and spin-strain coupling in the NV S=1 ground state}
\label{sec:Preliminaries}

\subsection{Employed coordinate systems}
In this work, we choose a cubic reference frame with crystal coordinates $XYZ$, where $\boldsymbol{e}_X = \left(1,0,0\right)^T$, $\boldsymbol{e}_Y = \left(0,1,0\right)^T$ and $\boldsymbol{e}_Z = \left( 0,0,1 \right)^T$ [see Tab.\,\ref{tab:CoordinateSystems}]. Since NV centers can have four different orientations in the diamond lattice, we further employ four NV reference frames $xyz_k$ with $k \in \{\mathrm{NV1}, \mathrm{NV2}, \mathrm{NV3}, \mathrm{NV4}\}$. Each $xyz_k$ is determined by a set of orthonormal basis vectors $\boldsymbol{e}_i^k$ with $i\in \{x,y,z\}$, which are defined in Tab.\,\ref{tab:CoordinateSystems}. Our choice of $xyz_k$ is such that the $z$-axes serve as the main symmetry axes of the four defect orientations, and the $y$-axes lie in NV symmetry planes. Defining the $xyz_k$ with the $x$-axes in the reflection planes is also common, but does not change the formalism we present in this work [see App.\,\ref{app:influence_NV_coordinate_systems} for a short comparison]. In the following, unless noted otherwise, we refer to the NV frame for orientation NV1.
\begin{table*}[!]
	\setlength{\tabcolsep}{5pt}
	\renewcommand{\arraystretch}{2.4}
	\centering
	\caption{Definition and graphical representation of crystal $(XYZ)$ and NV $(xyz_k)$ coordinate systems employed in this work (for clarity, $x$-axes are represented by dashed arrows, $y$-axes by dotted arrows and $z$-axes by solid arrows). The given rotations $\boldsymbol{K}_k$ with $k \in \{\mathrm{NV2}, \mathrm{NV3}, \mathrm{NV4}\}$ describe a coordinate transformation of NV1 into NV2-4. $\boldsymbol{L}_\mathrm{NV1} = \boldsymbol{R}_{[001]}(-3\pi/4) \boldsymbol{R}_{[\bar{1}10]}(-\alpha_\mathrm{NV})$ represents the coordinate system transformation of $XYZ\rightarrow xyz_\mathrm{NV1}$ with $\alpha_\mathrm{NV} = \arccos(1/\sqrt{3})$. To obtain the rotations $\boldsymbol{\tilde{K}}_k$ in Kelvin notation, we replace all $\boldsymbol{R}_{\boldsymbol{n}} (\theta)$ with $\boldsymbol{\tilde{R}}_{\boldsymbol{n}} (\theta)$ (see App.\,\ref{sec:app_definition_rotationmatrices} for definitions of rotation matrices in standard or Kelvin notation).}
	\begin{tabular}{p{5.2cm} c c c c c c}
		\hline
		\hline		
		\multirow{5}{*}{\includegraphics[width=5.2cm]{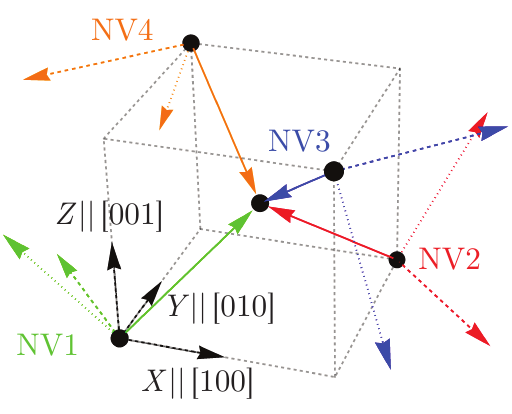}} & 	NV orientation & $ \sqrt{2}\boldsymbol{e}_x$ & $\sqrt{6}\boldsymbol{e}_y$ & $\sqrt{3}\boldsymbol{e}_z$ & $\boldsymbol{K}_k$ & $\boldsymbol{L}_k$ \\
		\cline{2-7}
		& NV1 &  $[\bar{1}10]$ & $[\bar{1}\bar{1}2]$ & $[111]$ &  $\mathds{1}$ & $\boldsymbol{L}_\mathrm{NV1}$\\
		& NV2 &	 $[1\bar{1}0]$ & $[112]$	&	$[\bar{1}\bar{1}1]$	&	$\boldsymbol{R}_{[001]}(\pi)$ & $\boldsymbol{L}_\mathrm{NV1} \boldsymbol{K}_\mathrm{NV2}$\\
		& NV3	&	$[110]$	&	$[1\bar{1}\bar{2}]$	&	$[\bar{1}1\bar{1}]$	& $\boldsymbol{R}_{[010]}(\pi)$ & $\boldsymbol{L}_\mathrm{NV1} \boldsymbol{K}_\mathrm{NV3}$\\ 
		&NV4	&	$[\bar{1}\bar{1}0]$	&	$[\bar{1}1\bar{2}]$	&	$[1\bar{1}\bar{1}]$	&		$\boldsymbol{R}_{[100]}(\pi)$ & $\boldsymbol{L}_\mathrm{NV1} \boldsymbol{K}_\mathrm{NV4}$ \\
		\hline  
		\hline                
	\end{tabular}
	\label{tab:CoordinateSystems}
	\renewcommand{\arraystretch}{1}
\end{table*}

\subsection{Spin-stress coupling expressed in crystal coordinates $XYZ$}
The NV center consists of a substitutional nitrogen atom and a neighboring vacancy. In its orbital ground state, the negatively charged NV center forms an $S=1$ spin system, with the spin sublevels $\ket{0}$, $\ket{-1}$ and $\ket{+1}$ being eigenstates of the spin operator $S_z$ along the NV symmetry axis $z$ (i.e. $S_z\ket{m_s}=m_s\ket{m_s}$). In the absence of symmetry breaking fields, the electronic spin states $\ket{\pm1}$ are degenerate and shifted from $\ket{0}$ by a zero-field splitting $D_0 = 2.87~$GHz. An external magnetic field $\boldsymbol{B} = \left(B_x, B_y, B_z\right)^T$ induces a Zeeman splitting between the $\ket{\pm1}$ spin states and in presence of this field only, the NV spin is described by the Hamiltonian
\begin{equation}
H_0/h = D_0 S_z^2 + \gamma_\mathrm{NV} \boldsymbol{B} \boldsymbol{S},
\label{eq:NVHamiltonian}
\end{equation}
where $\gamma_\mathrm{NV} = 2.8$\,MHz/G is the NV gyromagnetic ratio, $h$ is Planck's constant, and $\boldsymbol{S} = \left(S_x, S_y, S_z\right)^T$ is the vector of the $S=1$ spin matrices.

The coupling between crystal deformation and the NV spin can be explained by a stress-induced change to the spin-spin interaction, which arises from the distortion of the unpaired spin density \cite{Barson2017}. The most general, symmetry-allowed spin-stress coupling Hamiltonian reads $H_\sigma = H_{\sigma 0} + H_{\sigma 1} + H_{\sigma 2}$, with
\begin{subequations}
	\label{eq:StressHamiltonian_NV1}
\begin{align}
H_{\sigma0}/h &= M_z S_z^2 \\
H_{\sigma1}/h &= N_x \{S_x, S_z\} + N_y \{ S_y, S_z\} \\
H_{\sigma2}/h &= M_x \left( S_y^2 - S_x^2 \right) + M_y \{ S_x, S_y\}, 
\end{align}
\end{subequations}
where $\{S_i, S_j\} = \left( S_i S_j + S_j S_i\right)$ is the anticommutator and $M_{x,y,z}$ and $N_{x,y}$ are coupling amplitudes (see below)\cite{Udvarhelyi2017}. The term $H_{\sigma0}$ preserves the NV symmetry and shifts $\ket{\pm1}$ with respect to $\ket{0}$. In contrast, $H_{\sigma2}$ leads to a coupling of spin sublevels $\ket{\pm1}$, while $H_{\sigma1}$ only has non-zero matrix elements between $\ket{0}$ and either $\ket{-1}$ or $\ket{+1}$. The stress-induced level shifts and splittings depend on NV orientation and are characterized by the five coupling amplitudes \cite{Udvarhelyi2017}
\begin{subequations}
	\label{eq:StressCouplingAmplitudes_XYZ_NV1}
	\begin{align}
	M_{x}^\mathrm{NV1} & = b \left( 2 \sigma_{ZZ} - \sigma_{XX} - \sigma_{YY}\right) \notag \\
		& \hspace{2cm} + c \left( 2 \sigma_{XY} - \sigma_{YZ} - \sigma_{XZ}\right) \\
	M_{y}^\mathrm{NV1} & =  \sqrt{3} b \left( \sigma_{XX} - \sigma_{YY} \right) + \sqrt{3} c \left( \sigma_{YZ} - \sigma_{XZ}\right) \\
	M_{z}^\mathrm{NV1} & = a_1 \left( \sigma_{XX} + \sigma_{YY} + \sigma_{ZZ}\right)  \notag \\
		&\hspace{2cm} + 2 a_2 \left( \sigma_{YZ} + \sigma_{XZ} + \sigma_{XY}\right)\\
	N_{x}^\mathrm{NV1} & = d \left( 2 \sigma_{ZZ} - \sigma_{XX} - \sigma_{YY}\right)  \notag \\
		&\hspace{2cm} + e \left( 2 \sigma_{XY} - \sigma_{YZ} - \sigma_{XZ}\right) \\
	N_{y}^\mathrm{NV1} & =  \sqrt{3} d \left( \sigma_{XX} - \sigma_{YY} \right) + \sqrt{3} e \left( \sigma_{YZ} - \sigma_{XZ}\right) 
	\end{align}
\end{subequations}
(given for NV orientation NV1). They further depend on the spin-stress coupling constants $a_1, a_2, b, c, d, e$ and the stress tensor components $\sigma_{IJ}$. Note that Eq.\,\eqref{eq:StressCouplingAmplitudes_XYZ_NV1} is true for any stress tensor and therefore provides a powerful tool to predict the effect of spin-stress coupling.

Before we include the remaining NV orientations NV2-4 in our formalism, we first demonstrate how Hamiltonian\,\eqref{eq:StressHamiltonian_NV1} can be used to predict stress-induced level shifts in the $S=1$ ground state. 
To that end, we consider a scenario in which no external magnetic field $\boldsymbol{B}$ is applied. Under such conditions, the terms in $H_{\sigma 1}$ are far off resonance and can be neglected to first order, resulting in the stress-induced level shifts
\begin{align}
\label{eq:StessInducedLevelShifts_noBfield_generalExpression}
\Delta_{\ket{\pm1}} & = \left( E_{\ket{\pm 1}}(P) - E_{\ket{\pm 1}}(P=0) \right)/h \notag \\
					& = \left( M_z \pm \sqrt{M_x^2 +M_y^2}  \right), 
\end{align} 
where $E_{\ket{\pm 1}}(P)$ denote the energies of the new eigenstates in the $\ket{\pm1}$ manifold with applied stress of amplitude $P$.
For uniaxial stress acting along $\boldsymbol{e}_P$, the stress tensor components are 
\begin{equation}
\label{eq:StressTensorComponents_XYZ}
\sigma_{IJ} = P \cos( \sphericalangle \boldsymbol{e}_P \boldsymbol{e}_I) \cos(\sphericalangle \boldsymbol{e}_P \boldsymbol{e}_J), 
\end{equation}
where $\sphericalangle \boldsymbol{e}_P \boldsymbol{e}_I$ and $\sphericalangle \boldsymbol{e}_P \boldsymbol{e}_J$ denote the angles between the applied stress and the crystal axes $\boldsymbol{e}_{I}$ and $\boldsymbol{e}_{J}$, with $I,J \in \{X,Y,Z\}$ \cite{Hughes1967}. For example, stresses along the $[100]$, $[110]$ and $[111]$ directions result in the stress tensors
\begin{subequations}
	\label{eq:StressTensors_100110111}
\begin{align}
\boldsymbol{\sigma}_{XYZ}^{[100]} &= P \begin{pmatrix} 1 & 0 & 0 \\ 0 & 0 & 0 \\ 0 & 0 & 0 \end{pmatrix}\\
\boldsymbol{\sigma}_{XYZ}^{[110]} &= \frac{P}{2} \begin{pmatrix} 1 & 1 & 0 \\ 1 & 1 & 0 \\ 0 & 0 & 0 \end{pmatrix}\\
\boldsymbol{\sigma}_{XYZ}^{[111]} &= \frac{P}{3} \begin{pmatrix} 1 & 1 & 1 \\ 1 & 1 & 1 \\ 1 & 1 & 1 \end{pmatrix},
\end{align}
\end{subequations}
where the subscript indicates definition of the stress tensors in the crystal coordinates $XYZ$. Consequently, the resulting level shifts are
\begin{subequations}
	\label{eq:SpinStressLevelshifts_NV1}
	\begin{align}
	\Delta_{\ket{\pm1}}^{[100]}/P & = a_1 \pm 2 b \\
	\Delta_{\ket{\pm1}}^{[110]}/P & =  a_1 + a_2 \pm (b - c) \\
	\Delta_{\ket{\pm1}}^{[111]}/P & = a_1 + 2 a_2. 
	\end{align}
\end{subequations}
It becomes clear that the four coupling constants $a_1$, $a_2$, $b$ and $c$ are necessary to fully describe spin-stress coupling for vanishing $\boldsymbol{B}$. Stress along the $[100]$ direction shifts the energy of the $\ket{\pm1}$ manifold by $a_1 P$ with respect to $|0\rangle$, whereas stress along $[110]$ shifts it by $(a_1 + a_2)P$. Similarly, the stress-induced splitting in the $\ket{\pm1}$ manifold is $4 b P$ in the case of $[100]$ stress, but $2 (b-c) P$ for $[110]$ stress \cite{Hughes1967, Barson2017}.
At this point, it is important to realize that a global phase uncertainty for coupling constants $b$ and $c$ exists, which arises from the fact that stress-induced splittings in the $\ket{\pm1}$ manifold are given by $\left( M_x^2 + M_y^2 \right)^{1/2}$. Therefore, the expressions $\Delta_{\ket{\pm1}}^{[100]}/P =  a_1 \mp 2b$ or $\Delta_{\ket{\pm1}}^{[110]}/P =  a_1 + a_2 \pm (c - b)$ are also fully justified.  Here, compared to Eq.\,\eqref{eq:SpinStressLevelshifts_NV1}, the signs of $b$ and $c$ are flipped. To keep the formalism in this work as consistent as possible with existing literature \cite{Barson2017, Udvarhelyi2017}, we choose to work with the notation from Eq.\,\eqref{eq:SpinStressLevelshifts_NV1}.

So far, we considered the response of NV centers oriented as NV1 to stress described by the tensor $\boldsymbol{\sigma}_{XYZ} \equiv \boldsymbol{\sigma}_{XYZ}^\mathrm{NV1}$. To include the remaining three NV orientations, we express $\boldsymbol{\sigma}_{XYZ}^\mathrm{NV1}$ in the reference frames of NV2-4 by performing the coordinate system transformation
\begin{equation}
\label{eq:StressTensor_TransformationCoordinateSystem}
\boldsymbol{\sigma}_{XYZ}^k = \boldsymbol{K}_k \cdot \boldsymbol{\sigma}_{XYZ}^\mathrm{NV1} \cdot  \boldsymbol{K}_k^T
\end{equation}  
with $k \in \{\mathrm{NV2}, \mathrm{NV3}, \mathrm{NV4}\}$. The rotations $\boldsymbol{K}_k$ are given in Tab.\,\ref{tab:CoordinateSystems} [for a definition of the rotation matrices $\boldsymbol{R}_{\boldsymbol{n}} \left(\theta\right)$ see App.\,\ref{sec:app_definition_rotationmatrices}]. 
We then replace $\sigma_{IJ}$ in Eq.\,\eqref{eq:StressCouplingAmplitudes_XYZ_NV1} with the corresponding values from $\boldsymbol{\sigma}_{XYZ}^k$, thereby obtaining expressions for the coupling amplitudes of NV2-4. The stress-induced level shifts $\Delta_{\ket{\pm1}}$ for NV2-4 are obtained as described and are summarized in Tab.\,\ref{tab:Summary_LevelShifts_Stress_Strain}. For stress along the $[100]$ direction, all four NV orientations are affected in the same way. In contrast, NV orientations NV1+2 and NV3+4 exhibit different behavior for stress along $[110]$. Finally, in the case of stress along the $[111]$ direction, the symmetry of NV1 is preserved and it experiences level shifts only. This does not hold for NV2-4, which exhibit identical shifts and splittings \cite{Hughes1967, Rogers2015}.
Note that a Mathematica file is provided to reproduce our calculations in detail (see supplemental material).
\begin{table*}
	\renewcommand{\arraystretch}{1.71}
	\centering
	\caption{Overview over NV orientations NV1-4 and the corresponding level shifts $\Delta_{\ket{\pm1}}$ for $\boldsymbol{B} = 0$ and stresses along the $[100]$, $[110]$ and $[111]$ directions, expressed in terms of stress (third column) and strain (fourth column). $a_1$, $a_2$, $b$ and $c$ denote the spin-stress coupling constants while $A_1$, $A_2$, $B$ and $C$ represent the spin-strain coupling constants. We use $\gamma = (C_{11}+2C_{12})/C_{44}$, the Poisson ratio $\nu = C_{12}/(C_{11} + C_{12})$ and strain amplitude $\epsilon = P/E$, with $E = (C_{11} - C_{12})(C_{11} + 2 C_{12})/(C_{11} + C_{12})$ being the Young's modulus, to shorten the expressions for strain-induced level shifts. Also note that the strain-induced level shifts are expressed in the engineering strain framework. To convert to pure strain, use $\gamma = 2(C_{11}+2C_{12})/C_{44}$ [see App.\,\ref{sec:app_engineering_vs_pure_strain}]. $(C_{11}, C_{12}, C_{44}) = (1076, 125, 576)$\,GPa are the stiffness tensor components of diamond \cite{Kaxiras2003, Gross2012}.}
\begin{tabular}{p{3cm} c c c c c}
	\hline	
	\hline
	\centering{stress $\boldsymbol{P}$}	&	NV orientation	& &	$\Delta_{\ket{\pm1}}/P$ & & $\Delta_{\ket{\pm1}}/\epsilon$   \\
	\hline
	\multirow{4}{*}{\vspace{-0.2cm}\includegraphics{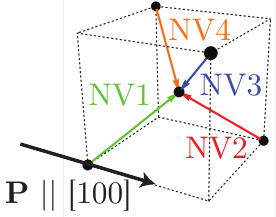}}	&	NV1	& \multirow{4}{*}[6.25ex]{\rdelim\}{3.85}{0.5cm}} & 	\multirow{4}{*}{$a_1 \pm 2b$} &  & 	\multirow{4}{*}{$\frac{(1-2\nu)}{3}(A_1 + 2 A_2) \pm  \frac{(2+2\nu)}{3}(B - \sqrt{2} C)$}	\\
	&	NV2	& & & &	\\
	&	NV3	& & & &\\
	&	NV4	& & & &\\
	\hline
	\multirow{4}{*}{\vspace{-0.2cm}\includegraphics{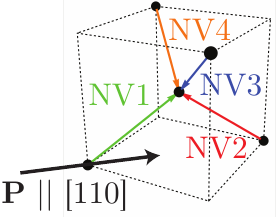}}	&	NV1	& \multirow{2}{*}[2ex]{\rdelim\}{2}{0.5cm}} & 	\multirow{2}{*}{$a_1+a_2  \pm (b-c)$} & \multirow{4}{*}{\hspace{0.3cm}} & \multirow{2}{*}{\shortstack{$\frac{1-2\nu}{6}( A_1 (2+\gamma) - A_2(\gamma-4))$ \\ $\pm \frac{1}{3} \left[ B(\gamma (1-2\nu)-(1+\nu)) + \frac{C}{\sqrt{2}}(\gamma (1 - 2\nu)+2(1+\nu)) \right]$}}	\\
	&	NV2 & & &	&\\
	&	NV3	& \multirow{2}{*}[2ex]{\rdelim\}{2}{0.5cm}} & 	\multirow{2}{*}{$a_1 - a_2  \pm (b+c)$} & \multirow{4}{*}{\hspace{0.3cm}} & \multirow{2}{*}{\shortstack{$\frac{1-2\nu}{6}( A_1 (2 - \gamma) + A_2 (\gamma + 4))$ \\ $\pm \frac{1}{3} \left[ B(\gamma (1-2\nu) + (1+\nu)) + \frac{C}{\sqrt{2}}(\gamma (1 - 2\nu) - 2(1+\nu)) \right]$}}	\\
	&	NV4	& & & & 	\\
	\hline
	\multirow{4}{*}{\vspace{-0.2cm}\includegraphics{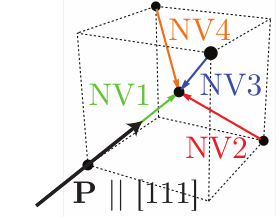}}	&	NV1	&  & 	$a_1 + 2 a_2$ & & $ \frac{1-2\nu}{3} ((A_1-A_2)\gamma + A_1 + 2A_2) $	\\
	&	NV2	& \multirow{2}{*}[1.3ex]{\rdelim\}{2.8}{0.5cm}}&  \multirow{3}{*}{$a_1-\frac{2}{3} a_2 \pm \frac{4}{3}c$} & &  \multirow{3}{*}{\shortstack{$\frac{1-2\nu}{9} (-(A_1-A_2)\gamma + 3(A_1 + 2A_2))$ \\ $\pm \frac{2\sqrt{2}}{9} \gamma (1-2\nu) (C+\sqrt{2}B)$}}	\\
	&	NV3	& & & &\\
	&	NV4	& & & &\\
	\hline  
	\hline                
\end{tabular}
\label{tab:Summary_LevelShifts_Stress_Strain}
\renewcommand{\arraystretch}{1}
\end{table*}

\subsection{Spin-strain coupling expressed in NV coordinate systems $xyz_k$}
Expressing the coupling between lattice deformation and NV spin in terms of stress defined in $XYZ$ is rather straightforward and leads to simple expressions for $\Delta_{\ket{\pm1}}$ [see Eq.\,\eqref{eq:SpinStressLevelshifts_NV1}]. Despite the simplicity of this approach, several past works employed a formalism based on strain defined in one of the NV coordinate systems $xyz_k$ \cite{Teissier2014, Teissier2017, Ovartchaiyapong2014, Lee2016}. To unify the two notations and allow for comparison of published results, we now show in detail how the spin-stress description is translated into the strain framework. First, we convert the spin-stress amplitudes of NV1 [see Eq.\,\eqref{eq:StressCouplingAmplitudes_XYZ_NV1}]  into spin-strain coupling amplitudes by expressing the stress tensor components $\sigma_{IJ}$ in terms of strain tensor components $\epsilon_{ij}$, which are now defined in the $xyz$ coordinate system of NV1. We then include the NV orientations NV2-4 by transforming $\boldsymbol{\epsilon}_{xyz}$ into the corresponding reference frames.

To find expressions for the spin-strain coupling amplitudes of NV1, we first link the stress tensor $\boldsymbol{\sigma}_{XYZ}^\mathrm{NV1}$ to the strain tensor $\boldsymbol{\epsilon}_{xyz}^\mathrm{NV1}$ by \cite{Barson2017}
\begin{equation}
	\boldsymbol{\tilde{\sigma}}_{XYZ}^\mathrm{NV1} = \boldsymbol{\tilde{C}}_{XYZ} \boldsymbol{\tilde{L}}_\mathrm{NV1}^T \boldsymbol{\tilde{\epsilon}}_{xyz}^\mathrm{NV1}.
\label{eq:StressXYZ2strainxyzNV1}
\end{equation}
Here, $\boldsymbol{\tilde{L}}_\mathrm{NV1} =  \boldsymbol{\tilde{R}}_{[001]}(-3\pi/4) \boldsymbol{\tilde{R}}_{[\bar{1}10]}(-\alpha_\mathrm{NV})$ describes the coordinate system transformation from $XYZ$ to $xyz_\mathrm{NV1}$ [see Tab.\,\ref{tab:CoordinateSystems} and App.\,\ref{sec:app_definition_rotationmatrices} for definition of rotation matrices]. As indicated by $\boldsymbol{\tilde{L}}$ and $\boldsymbol{\tilde{R}}$, we express the transformation matrices $\boldsymbol{L}$ and $\boldsymbol{R}$ in Kelvin notation to write Hooke's law in vectorial form, where $\boldsymbol{\tilde{\sigma}}$ and $\boldsymbol{\tilde{\epsilon}}$ are $6\times 1$ vectors and the stiffness tensor $\boldsymbol{\tilde{C}}$ is a $6\times6$ matrix [see App.\,\ref{sec:app_kelvinnotation}] \cite{Mehrabadi1995, Nagel2016}.
We then replace $\sigma_{IJ}$ in Eq.\,\eqref{eq:StressCouplingAmplitudes_XYZ_NV1} with the result from Eq.\,\eqref{eq:StressXYZ2strainxyzNV1} and obtain the NV1 spin-strain coupling amplitudes
\begin{subequations}
	\label{eq:StrainCouplingAmplitudes_xyz_NV1}
	\begin{align}
	M_{x}^\mathrm{NV1} & =  B \left(\epsilon_{xx} - \epsilon_{yy} \right) + 2 C \epsilon_{yz}\\
	M_{y}^\mathrm{NV1} & = -2 B \epsilon_{xy} - 2 C \epsilon_{xz} \\
	M_{z}^\mathrm{NV1} & = A_1 \epsilon_{zz} + A_2  \left(\epsilon_{xx} + \epsilon_{yy} \right)\\
	N_{x}^\mathrm{NV1} & = D \left(\epsilon_{xx} - \epsilon_{yy} \right) + 2 E \epsilon_{yz}\\
	N_{y}^\mathrm{NV1} & =   -2 D \epsilon_{xy} - 2 E \epsilon_{xz},
	\end{align}
\end{subequations}
which depend on the spin-strain coupling constants
\begin{subequations}
	\label{eq:StrainCouplingConstants_LudlowForm}
	\begin{align}
	A_1 & = a_1 (C_{11} + 2C_{12}) + 4 a_2 C_{44}  \\ 
	A_2 & = a_1 (C_{11} + 2C_{12}) -  2 a_2 C_{44} \\ 
	B & = -b(C_{11} - C_{12}) -  2 c C_{44}  \\ 
	C & = \sqrt{2} b  (C_{11} - C_{12}) -  \sqrt{2} c C_{44} \\
	D & = -d(C_{11} - C_{12}) -  2 e C_{44}  \\ 
	E & = \sqrt{2} d  (C_{11} - C_{12}) - \sqrt{2} e C_{44}.  
	\end{align}
\end{subequations}
As our formalism relies on the engineering strain convention, the relations in Eq.\,\eqref{eq:StrainCouplingConstants_LudlowForm} differ by a factor of 2 in the $C_{44}$ terms compared to other work\cite{Ludlow1968, Barson2017}, where the pure strain convention is used [see App.\,\ref{sec:app_engineering_vs_pure_strain} for a brief explanation].

As is evident from Eq.\,\eqref{eq:StressXYZ2strainxyzNV1}, the form of the spin-strain coupling amplitudes in Eq.\,\eqref{eq:StrainCouplingAmplitudes_xyz_NV1} for the same NV orientation depends on the employed NV reference frame. Working in the spin-strain framework therefore requires great care when for example strain-induced level shifts are to be calculated or spin-strain coupling constants need to be determined. We show in App.\,\ref{sec:app_definitionxyz_couplingamplitudes} that confusion can be avoided, even if different definitions of NV reference frames for the same NV orientation are employed, if the strain tensor components $\epsilon_{ij}$ in Eq.\,\eqref{eq:StrainCouplingAmplitudes_xyz_NV1} are expressed in terms of the original stress tensor $\boldsymbol{\sigma}_{XYZ}^\mathrm{NV1}$ by inverting Eq.\,\eqref{eq:StressXYZ2strainxyzNV1}. 

For uniaxial stresses along the $[100]$, $[110]$ and $[111]$ directions [see Eq.\,\eqref{eq:StressTensors_100110111}], the strain tensors in the NV1 reference frame obtained from Eq.\,\eqref{eq:StressXYZ2strainxyzNV1} are
\begin{widetext}
\begin{subequations}
	\label{eq:straintensorsNV1_100110111}
	\begin{align}
	\boldsymbol{\epsilon}_{xyz}^{[100]} & = \epsilon \begin{pmatrix}  \frac{1-\nu}{2}  & \frac{1+\nu}{\sqrt{12}} & -\frac{1+\nu}{\sqrt{6}} \\ \frac{1+\nu}{\sqrt{12}} & \frac{1-5\nu}{6} & -\frac{1+\nu}{\sqrt{18}}\\ -\frac{1+\nu}{\sqrt{6}} & -\frac{1+\nu}{\sqrt{18}} & \frac{1-2\nu}{3}  \end{pmatrix} \\	
	\boldsymbol{\epsilon}_{xyz}^{[110]} & = \epsilon \begin{pmatrix} \frac{2-2\nu-\gamma(1-2\nu)}{4}  & 0 & 0 \\ 0 & \frac{2-10\nu + \gamma(1-2\nu)}{12} & \frac{-2-2\nu -\gamma(1-2\nu)}{72} \\ 0 & \frac{-2-2\nu -\gamma(1-2\nu)}{72} & \frac{2(1-2\nu)+\gamma(1-2\nu)}{6} \end{pmatrix}\\
	\boldsymbol{\epsilon}_{xyz}^{[111]} & = \epsilon \begin{pmatrix} \frac{2(1-2\nu)-\gamma(1-2\nu)}{6} & 0 & 0 \\ 0 & \frac{2(1-2\nu)-\gamma(1-2\nu)}{6} & 0 \\ 0 & 0 & \frac{1-2\nu + \gamma(1 -2\nu)}{3} \end{pmatrix},  
	\end{align} 
\end{subequations}
\end{widetext}
where we introduced $\gamma = (C_{11}+2C_{12})/C_{44}$, the Poisson ratio $\nu = C_{12}/(C_{11} + C_{12})$ and the strain amplitude $\epsilon = P/E$ to shorten the notation. $P$ is the applied stress and $E = (C_{11} - C_{12})(C_{11} + 2 C_{12})/(C_{11} + C_{12})$ is the Young's modulus \cite{Gross2012}. The associated level shifts for $B=0$, which we obtain by combining Eqs.\,\eqref{eq:StessInducedLevelShifts_noBfield_generalExpression}, \,\eqref{eq:StrainCouplingAmplitudes_xyz_NV1} and \eqref{eq:straintensorsNV1_100110111}, are 
\begin{subequations}
	\label{eq:levelshifts100110111_strain_NV1}
	\begin{align}
	\Delta_{|\pm1\rangle}^{[100]}/\epsilon & = \frac{(1-2\nu)}{3}(A_1 + 2 A_2) \notag \\
	&\,\,\,\,\, \pm  \frac{(2 + 2\nu)}{3}(B - \sqrt{2} C)  \\
	\Delta_{|\pm1\rangle}^{[110]}/\epsilon & = \frac{1-2\nu}{6}( A_1 (2+\gamma) - A_2(\gamma-4)) \notag \\
	&\,\,\,\,\, \pm \frac{1}{3} [ B(\gamma (1-2\nu)-(1+\nu)) \notag \\
	&\,\,\,\,\, \,\,\,\,\, + \frac{C}{\sqrt{2}}(\gamma (1 - 2\nu)+2(1+\nu)) ] \\
	\Delta_{|\pm1\rangle}^{[111]}/\epsilon & = \frac{1-2\nu}{3} ((A_1-A_2)\gamma + A_1 + 2A_2).	\end{align}
\end{subequations}

To find the strain-induced level shifts for all four NV orientations, we follow a similar approach as in the stress framework. We first use the relation
\begin{equation}
\boldsymbol{\tilde{\epsilon}}_{xyz}^k = \boldsymbol{\tilde{L}}_k \boldsymbol{\tilde{C}}_{XYZ}^{-1} \boldsymbol{\tilde{\sigma}}_{XYZ}^\mathrm{NV1}
\label{eq:Strainxyz2StressXYZNVk}
\end{equation}
with the rotations $\boldsymbol{\tilde{L}}_k$ as defined in Tab.\,\ref{tab:CoordinateSystems} to express stress defined in $XYZ$ in terms of strain reference frames of NV2-4. By replacing the strain tensor components $\epsilon_{ij}^\mathrm{NV1}$ in Eq.\,\eqref{eq:StrainCouplingAmplitudes_xyz_NV1} with the $\epsilon_{ij}^k$ from Eq.\,\eqref{eq:Strainxyz2StressXYZNVk}, we then obtain the strain coupling amplitudes for NV2-4. The resulting level shifts are summarized in Tab.\,\ref{tab:Summary_LevelShifts_Stress_Strain}. Obviously, the expressions for the stress-induced level shifts are much more compact. This observation, together with the fact that all strain tensor components $\epsilon_{ij}^k$ should be expressed in terms of the original stress tensor $\sigma_{IJ}$ to avoid confusion regarding the actual definition of the $k$th NV reference frame, supports the notion that the stress formalism is much more effective in describing NV spin-oscillator coupling than strain.
Before we continue, we also want to point out that the presented formalism not only applies to the $S=1$ ground state manifold of the NV center, but can also be used to describe the influence of crystal deformation on the NV's $S=1$ excited state \cite{Lee2016} and the NV's $S=0$ ground state leves \cite{Rogers2015}, where the orbital symmetries of the involved states are identical. 

\section{Stress and strain in cantilevers}
\label{sec:stress_cantilevers}
To illustrate how the coupling formalism developed here can be applied, we will now derive an expression for the stress tensor in singly-clamped cantilever beams that are bent by a static external force $\boldsymbol{V}$. Such beams are currently the most common choice if spin-stress coupling in diamond-based hybrid spin-oscillator systems is to be quantified \cite{Teissier2014, Ovartchaiyapong2014, Barson2017}, since the occurring stress can be described analytically using a relatively simple approach.

\subsection{Cantilever coordinate system $\tilde{x}\tilde{y}\tilde{z}$ and sign conventions}
\label{subsec:cantilevercoordinates_signconventions}
\begin{figure}[h]
	\centering
	\includegraphics[]{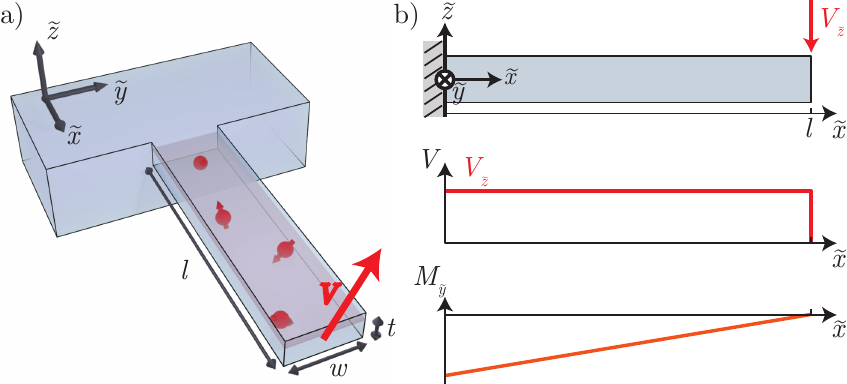}
	\caption{Beam bending with a transverse force. 
		a) A singly-clamped cantilever with dimensions width $w$, length $l$ and thickness $t$ is subject to an external shear force $\boldsymbol{V} = (V_{\tilde{x}}, V_{\tilde{y}}, V_{\tilde{z}})^T$, which is applied at the tip of the cantilever, i.e. at $\tilde{x} = l$.
		b) A positive shear force of amplitude $V_{\tilde{z}}$ pushes the cantilever downwards and induces a negative bending moment $M_{\tilde{y}}$. As the induced shear force remains constant along the beam, the bending moment decreases linearly from tip to root.}
	\label{fig:beambending_transverseforce}
\end{figure}

We begin our discussion by defining sign conventions for shear force, bending moment, coordinate directions, beam deflection, lateral forces and strain or stress \cite{Fenner2012}. 
In general we consider a cantilever of length $l$, which has a rectangular cross section of width $w$ and thickness $t$ with $l \gg w,t$. The cantilever coordinate system $\tilde{x}\tilde{y}\tilde{z}$ is chosen such that length $l$ is defined along $\boldsymbol{e}_{\tilde{x}}$,  width $w$ along $\boldsymbol{e}_{\tilde{y}}$ and thickness $t$ along $\boldsymbol{e}_{\tilde{z}}$ (see Fig.\,\ref{fig:beambending_transverseforce}a)). 
The $\tilde{x}$ axis has its origin at the clamped end of the beam and $\tilde{y}$ and $\tilde{z}$ are defined with respect to the cross section's centroid. Points that lie within the beam are therefore described by $\tilde{x} \in [0,l]$, $\tilde{y} \in [-w/2, w/2]$ and $\tilde{z} \in [-t/2 , t/2]$.
Lateral deflection $u$ is chosen to be positive along $-\boldsymbol{e}_{\tilde{z}}$. 
Shear forces $\boldsymbol{V}$ are defined positive if they cause the beam to rotate clockwise. For example, an external force pointing along $ -\boldsymbol{e}_{\tilde{z}}$ and applied at positive $\tilde{x}$ would rotate the beam clockwise about the $\tilde{y}$ axis when looking along $\boldsymbol{e}_{\tilde{y}}$ and is therefore considered positive.
Induced bending moments are defined to be positive if they correspond to a sagging behavior of the beam, while negative bending moments refer to hogging [scenario in Fig.\,\ref{fig:beambending_transverseforce}b].  
Finally, tension (compression) relates to positive (negative) strain and stress amplitudes.

\subsection{Stress tensor of a singly-clamped cantilever}
To derive the stress tensor in a singly-clamped cantilever, we consider a transverse force of magnitude $V_{\tilde{z}}$ that is applied at the tip of the beam and pushes it in the $-\boldsymbol{e}_z$ direction. This force induces the negative bending moment
\begin{equation}
M_{\tilde{y}}(\tilde{x}) = -(l-\tilde{x}) V_{\tilde{z}}
\label{eq:bendingmoment_transverseshear}
\end{equation}
for $\tilde{x} \in [0,l]$ and causes normal stress that points along $\boldsymbol{e}_{\tilde{x}}$. For a true beam, which satisfies $l \gg w,t$, we can apply the flexural formula \cite{Shames1992, Fenner2012, Cleland2003} to find the induced normal stress
\begin{equation}
\sigma_{n}(\tilde{x},\tilde{z}) = -\frac{\tilde{z}}{I_{\tilde{z}}}M_{\tilde{y}}(\tilde{x}) = \frac{\tilde{z} (l-\tilde{x})}{I_{\tilde{z}}}V_{\tilde{z}}
\label{eq:stress_flexformula}
\end{equation}
with the moment of inertia $I_{\tilde{z}} = w t^3/12$. Bending the cantilever downwards induces tensile stress in the top half of the beam $(\tilde{z}>0)$ and compressive stress in the lower half  $(\tilde{z}<0)$. 
Moreover, $\sigma_{n}(\tilde{x},\tilde{z})$ decreases linearly from root to tip and from the neutral plane, i.e. the plane with $\tilde{z} = 0$, towards top and bottom surfaces at $\tilde{z} = \pm t/2$. In addition to normal stress, the applied transverse force also gives rise to shear stress
\begin{equation}
	\sigma_{s}(\tilde{z}) = \frac{V_{\tilde{z}}}{2 I_{\tilde{z}}} \left[\left(\frac{t}{2}\right)^2 - \tilde{z}^2\right]
\label{eq:shearstress}
\end{equation}
in the cantilever \cite{Fenner2012}, which vanishes at the top and bottom surfaces and is maximized in the beam's neutral plane.  

To link this discussion to the spin-stress coupling amplitudes in the previous section [see e.g.\,Eq.\,\eqref{eq:StressCouplingAmplitudes_XYZ_NV1}], we now formulate a stress tensor that corresponds to the influence of an external shear force $\boldsymbol{V} = \left( V_{\tilde{x}},V_{\tilde{y}},V_{\tilde{z}}\right)^T$ applied to a cantilever as shown in Fig.\,\ref{fig:beambending_transverseforce}a). 
In our example, the cantilever is oriented such that its coordinate system $\tilde{x} \tilde{y} \tilde{z}$ coincides with the crystal coordinate system $XYZ$.
The axial force component $V_{\tilde{x}}$ causes a constant normal stress $\sigma_{XX} = V_{\tilde{x}}/A$ along the beam where $A = w t$ is the beam's cross-sectional area. 
As we know from Eqs.\,\eqref{eq:stress_flexformula} and \eqref{eq:shearstress}, the transverse component $V_{\tilde{z}} \parallel -\boldsymbol{e}_Z$ induces normal stress $\sigma_{XX} = V_{\tilde{z}} \tilde{z} (l-\tilde{x})/I_{\tilde{z}}$ as well as shear stresses $\sigma_{XZ} = \sigma_{ZX} = V_{\tilde{z}} \left[(t/2)^2 - \tilde{z}^2 \right]/2 I_{\tilde{z}}$. 
In analogy, the transverse component $V_{\tilde{y}} \parallel +\boldsymbol{e}_Y$ causes normal stress $\sigma_{XX} = V_{\tilde{y}} \tilde{y} (l-\tilde{x})/I_{\tilde{y}}$ and shear stresses $\sigma_{XY} = \sigma_{YX} = V_{\tilde{y}} \left[(w/2)^2 - \tilde{y}^2 \right]/2 I_{\tilde{y}}$ with $I_{\tilde{y}} = t w^3/12$. 
All in all, the final stress tensor for small cantilever deflections is
\begin{widetext}
	\begin{equation}
	\boldsymbol{\sigma}_{XYZ}^c =  \begin{pmatrix} 
	\frac{V_{\tilde{x}}}{wt} + (l-\tilde{x})\left(\frac{\tilde{z} V_{\tilde{z}}}{I_{\tilde{z}}} + \frac{\tilde{y} V_{\tilde{y}}}{I_{\tilde{y}}}\right) & \frac{ V_{\tilde{y}}}{2 I_{\tilde{y}}} \left[ \left(\frac{w}{2}\right)^2 - \tilde{y}^2 \right] & \frac{ V_{\tilde{z}}}{2 I_{\tilde{z}}} \left[ \left(\frac{t}{2}\right)^2 - \tilde{z}^2 \right]\\ 
	\frac{ V_{\tilde{y}}}{2 I_{\tilde{y}}} \left[ \left(\frac{w}{2}\right)^2 - \tilde{y}^2 \right] & 0 & 0 \\ 
	\frac{ V_{\tilde{z}}}{2 I_{\tilde{z}}} \left[ \left(\frac{t}{2}\right)^2 - \tilde{z}^2 \right] & 0 & 0 
	\end{pmatrix}.
	\label{eq:stresstensor_cantileverbending}
	\end{equation}
 \end{widetext}
Under the assumption of a purely transverse force along $-\boldsymbol{e}_{\tilde{z}}$, the stress tensor close to the beam's top surface simplifies to 
\begin{equation}
\boldsymbol{\sigma}_{XYZ}^c = P(\tilde{x},\tilde{z}) \begin{pmatrix} 
1 & 0 & 0 \\ 
0 & 0 & 0 \\ 
0 & 0 & 0 
\end{pmatrix}
\label{eq:stresstensor_cantileverbending_simplified}
\end{equation}
with $P(\tilde{x},\tilde{z}) = \frac{\tilde{z}V_{\tilde{z}}}{I_{\tilde{z}}} (l-\tilde{x})$ being the applied stress amplitude. We  point out that pushing along $-\boldsymbol{e}_{\tilde{z}}$ on a cantilever oriented along the $[100]$ direction introduces uniaxial stress along the $[100]$ direction. Consequently, stress in cantilevers of different orientations can be obtained by making an appropriate coordinate system transformation.

From an experimental point of view, it is often desirable to express $P(\tilde{x},\tilde{z})$ in terms of the induced cantilever deflection $u$.
From Euler-Bernoulli beam theory we know that a force $V_{\tilde{z}}$ applied at the beam's end causes a beam deflection $u(\tilde{x})$ of the form \cite{Cleland2003}
\begin{equation}
u(\tilde{x}) = \frac{V_{\tilde{z}}}{E I_{\tilde{z}}}\left( \frac{l \tilde{x}^2}{2} - \frac{\tilde{x}^3}{6}\right).
\label{eq:beamdeflection}
\end{equation}
We can thus link the applied force $V_{\tilde{z}}$ to the maximum beam displacement $u(l)$ via the expression
\begin{equation}
V_{\tilde{z}} = \frac{3 E I_{\tilde{z}}}{l^3} u(l), 
\label{eq:force_deflection}
\end{equation}
and the stress amplitude $P(\tilde{x},\tilde{z})$ becomes
\begin{equation}
P(\tilde{x},\tilde{z})= \frac{3 \tilde{z} E}{l^3}(l-\tilde{x}) u(l)
\label{eq:HybridSystem_CantileverPhysics_BeamDeflection_stressDeflection}
\end{equation}
where $u(l)$ now represents the cantilever deflection measured at $\tilde{x} = l$ and $E$ is the Young's modulus.

\section{Determining spin-stress coupling constants in diamond-based hybrid systems}
\label{sec:determining_spinstresscouplingconstants}
After establishing a full and consistent treatment of spin-stress coupling in the NV ground state and deriving an expression for the stress tensor in a singly-clamped cantilever, we now present bending experiments which we performed to characterize the spin-stress and spin-strain coupling constants. In our original analysis of these measurements by Teissier \textit{et al.}\cite{Teissier2014}, we used an oversimplified theoretical description of the coupling mechanism, which neglected shear strain and the Poisson effect. With the formalism developed here, we can now extract the correct spin-stress coupling constants and compare them to existing literature.

The diamond cantilevers investigated by Teissier \textit{et al.} were aligned such that $\boldsymbol{e}_{\tilde{x}} \parallel [110]$ and $\boldsymbol{e}_{\tilde{z}} \parallel [001]$, had dimensions of $(w \times l \times t) = (3.5 \times 10-50 \times 0.2-1)\,\mu\text{m}^3$ and contained shallow implanted NV centers, which were located $\sim 17$\,nm below the top surface. We used a metal tip, placed at $\tilde{x}=l$, to displace the cantilever along $\boldsymbol{e}_{\tilde{z}}$. The resulting stress tensor reads
\begin{equation}
	\boldsymbol{\sigma}_{XYZ}^{[110]} = \frac{P}{2} \begin{pmatrix} 1 & 1 & 0 \\ 1 & 1 & 0 \\ 0 & 0 & 0 \end{pmatrix},
\label{eq:stresstensor_cantilever110}
\end{equation}
where the stress amplitude for shallow NV centers ($\tilde{z} \approx t/2$) located at the cantilever's base ($\tilde{x} \approx 0$) is given by [see Eq.\,\eqref{eq:HybridSystem_CantileverPhysics_BeamDeflection_stressDeflection}]
\begin{equation}
P \equiv P(0,t/2) = \frac{3}{2} \frac{t}{l^2} E u.
\label{eq:stressamplitude_cantilever110_shallowNVatbase}
\end{equation}
Since the metallic tip (tungsten) was about three orders of magnitude stiffer than the cantilever, the beam deflection $u$ was directly given by the piezo displacement amplitude of the tip.
 
For stress along the $[110]$ direction, the four possible NV orientations can be grouped into two subgroups with respect to their stress-induced level shifts and splittings: NV1+2 and NV3+4, which we will refer to as NVA and NVB in the following. The associated level shifts are (see Tab.\,\ref{tab:Summary_LevelShifts_Stress_Strain})
\begin{subequations}
	\label{eq:stress110_levelshiftsNVANVB}
	\begin{align}
	\Delta_{|\pm1\rangle}^\mathrm{NVA}/P  = (a_1 + a_2)\pm ( b- c) \\
	\Delta_{|\pm1\rangle}^\mathrm{NVB}/P  = (a_1 - a_2) \pm (b + c).
	\end{align}
\end{subequations}
To unambiguously identify all four spin-stress coupling constants, we investigated the stress-induced level shifts for both NV orientations by performing optically detected electron spin resonance (ESR) measurements [see App.\,\ref{sec:app_determination_spinstresscouplingconstants} for the analyzed ESR data sets] \cite{Teissier2014}. We then extracted spin sublevel shifts $\Delta_\parallel^{\mathrm{NVA},\mathrm{NVB}}$ and splittings $\Delta_\perp^{\mathrm{NVA},\mathrm{NVB}}$ as
\begin{subequations}
	\label{eq:stress110_shiftssplittings_NVANVB}
	\begin{align}
	\Delta_\parallel^\mathrm{NVA} P \equiv (\Delta_{|+1\rangle}^\mathrm{NVA} + \Delta_{|-1\rangle}^\mathrm{NVA})/2 & = (a_1 + a_2) P  \\
	\Delta_\perp^\mathrm{NVA} P \equiv (\Delta_{|+1\rangle}^\mathrm{NVA} - \Delta_{|-1\rangle}^\mathrm{NVA})/2 & = (b - c)P \\
	\Delta_\parallel^\mathrm{NVB} P \equiv (\Delta_{|+1\rangle}^\mathrm{NVB} + \Delta_{|-1\rangle}^\mathrm{NVB})/2 & = (a_1 - a_2) P  \\
	\Delta_\perp^\mathrm{NVB} P \equiv (\Delta_{|+1\rangle}^\mathrm{NVB} - \Delta_{|-1\rangle}^\mathrm{NVB})/2 & = (b + c) P.
	\end{align}
\end{subequations}
Finally, the spin-stress coupling constants are given by
\begin{subequations}
	\label{eq:stress110_spinstresscouplingconstants_levelshiftssplittingsNVANVB}
	\begin{align}
	a_1 & = (\Delta_\parallel^\mathrm{NVA} + \Delta_\parallel^\mathrm{NVB})/2   \\ 
	a_2 & = (\Delta_\parallel^\mathrm{NVA} - \Delta_\parallel^\mathrm{NVB})/2   \\ 
	b & = (\Delta_\perp^\mathrm{NVA} + \Delta_\perp^\mathrm{NVB})/2   \\ 
	c & = (\Delta_\perp^\mathrm{NVB} - \Delta_\perp^\mathrm{NVA})/2. 
	\end{align}
\end{subequations}

\begin{figure}
\centering
\includegraphics[]{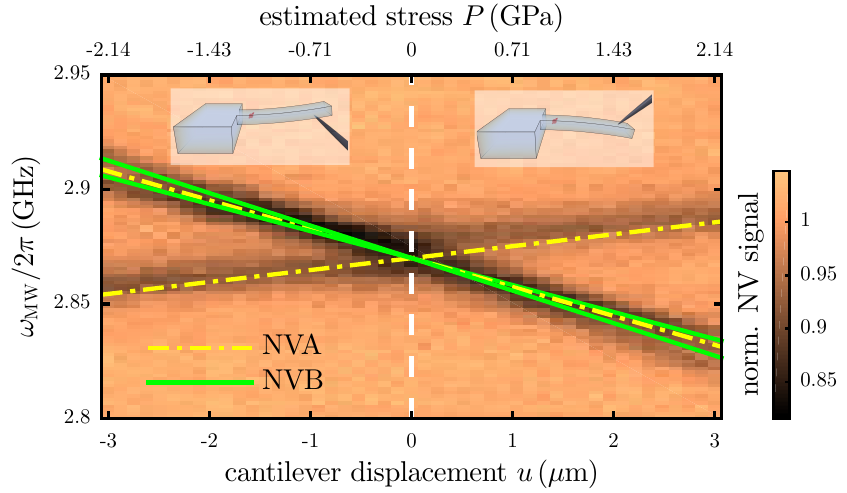}
\caption{Comparing experimental data with expected stress-induced level shifts (yellow dashed-dotted lines denote predictions for NV family NVA and green solid lines represent NV orientation NVB) yields a very good agreement between predicted spin-stress coupling parameters and our experiment. 
}
\label{fig:plotexpdatapredition}
\end{figure}
We measured $\Delta_\parallel^\mathrm{NVA,NVB}$ and $\Delta_\perp^\mathrm{NVA,NVB}$ for a total of five NV centers [three from orientation NVA and two from NVB, see App.\,\ref{sec:app_determination_spinstresscouplingconstants}) and determined the values
\begin{subequations}
	\label{eq:stress110_spinstresscouplingconstants_levelshiftssplittingsNVANVB_values}
	\begin{align}
	a_1 & = (-11.7 \pm 3.2)\text{\,MHz/GPa}  \\ 
	a_2 & = (6.5 \pm 3.2)\text{\,MHz/GPa}   \\ 
	b & = (7.1 \pm 0.8)\text{\,MHz/GPa}  \\ 
	c & = (-5.4 \pm 0.8)\text{\,MHz/GPa}. 
	\end{align}
\end{subequations}
for the spin-stress coupling constants. 
The given errors denote 68\,\% confidence intervals. They are rather large as the small number of NV centers we analyzed in our experiments was not sufficient to deal with systematic errors induced by e.g.\,different environmental stress fields resulting from crystal defects, surface roughness or the proximity of cantilever edges. Despite the large uncertainties, we find a very good agreement between the theoretically expected level shifts based on the spin-stress coupling constants in Eq.\,\eqref{eq:stress110_spinstresscouplingconstants_levelshiftssplittingsNVANVB_values} and typical experimental data shown in Fig.\,\ref{fig:plotexpdatapredition}. Finally, the spin-strain coupling constants $A_1, A_2, B, C$ 
\begin{subequations}
	\begin{align}
	\label{eq:StrainCouplingSingleSpin_BendingExperiments_Staticbending_straincouplingconstants_values}
	A_1 & = (-0.5 \pm 8.6)\text{\,GHz/strain} \\ 
	A_2 & = (-9.2 \pm 5.7)\text{\,GHz/strain}   \\ 
	B & = (-0.5 \pm 1.2)\text{\,GHz/strain}  \\ 
	C & = (14.0 \pm 1.3)\text{\,GHz/strain}. 
	\end{align}
\end{subequations}
are obtained via Eq.\,\eqref{eq:StrainCouplingConstants_LudlowForm}.
 
\begin{table}[]
 	\renewcommand{\arraystretch}{1.5}
 	\centering
 	\caption{Comparing spin-stress coupling constants from \cite{Teissier2014} and \cite{Barson2017}. The two sets of values differ by a factor of $\sim 2-3$ as well as in their signs, which can be explained by a potentially imprecise determination of cantilever dimensions in \cite{Teissier2014} and different sign conventions for tensile/compressive stress.}
 	\begin{tabular}{c c c}
 		\hline	
 		\hline
 		 &	This work	&	Barson\,\textit{et al.} \cite{Barson2017} \\
 		& MHz/GPa & MHz/GPa \\
 		\hline
 		$a_1$		& $-11.7 \pm 3.2$		& $4.86 \pm 0.02$	\\
 		$a_2$		& $6.5 \pm 3.2$			& $-3.7 \pm 0.2$	\\
 		$b$ 		& $7.1 \pm 0.8$			& $-2.3 \pm 0.3$	\\
 		$c$ 		& $-5.4 \pm 0.8$		& $3.5 \pm 0.3$	\\	
 		\hline  
 		\hline                
 	\end{tabular}
 	\label{tab:comparison_spinstresscouplingconstants}
 	\renewcommand{\arraystretch}{1}
 \end{table}
Similar values for the stress coupling constants, obtained through applying uniaxial stress to a diamond cube in a diamond anvil cell, were reported recently  \cite{Barson2017}. Comparing the two sets of values (see Tab.\,\ref{tab:comparison_spinstresscouplingconstants}) shows that both experiments find spin-stress coupling constants on the order of a few MHz/GPa. Yet they differ by a factor $\sim 2-3$ and in their signs. 
The origin of the sign discrepancy lies in different sign conventions for the applied stress. In Barson \textit{et al.}\cite{Barson2017}, compressive stress is defined to have positive amplitdues and causes the NV zero-field splitting $D_\mathrm{0}$ to increase. In our analysis, however, compressive stress is negative and increases $D_\mathrm{0}$ [see Fig.\,\ref{fig:plotexpdatapredition}]. Consequently, the spin-stress coupling constants have different signs.
We tentatively assign the mismatch in amplitude to uncertainties in the calibration of the applied stress with respect to cantilever deflection, which could be caused by imprecisely determined cantilever dimensions or inhomogeneous stress fields across the resonator. 
Note that uncertainties in $l$ pose a serious problem as the applied stress amplitude $P$ is proportional to $l^{-2}$. A potential measurement error in $l$ of $25$\,\% would result in values for stress coupling parameters almost identical in amplitude to the values reported by Barson \textit{et al.}\cite{Barson2017}. We thus suggest similar experiments to be conducted on better-defined geometries to reduce the uncertainty in cantilever dimensions.

\section{Summary and Outlook}
To summarize, we give a complete description of how the coupling between lattice deformation and the spin-degree of freedom of all four possible NV orientations can be described in terms of stress, defined in the crystal coordinate system $XYZ$, and strain, defined in the individual NV reference frames. We find that the stress-based approach is straightforward and yields compact expressions for stress-induced level shifts and splittings. In contrast, the strain-based formalism yields complicated expressions and requires extra care in the definition of the underlying coordinate reference frames. Since these are in general not given in literature, we suggest that future publications in the field employ the spin-stress formalism, such that published results can be compared more straightforwardly.
We further illustrate how the spin-stress formalism can be used to determine the spin-stress coupling constants. To that end, we derive the stress tensor in a singly-clamped cantilever and use it to reanalyze measurements from Teissier \textit{et al.} \cite{Teissier2014}. We find that the extracted values for the spin-stress coupling constants are in good agreement with other values reported in literature, but also point out differences in employed sign conventions.

With a correct and consistent framework at hand, we can now complete the characterization of spin-stress coupling in the NV $S=1$ ground state by quantifying the remaining spin-stress coupling constants $d$ and $e$. In our presented experiments, this was not possible as the relevant coupling $H_{\sigma 1}$ between $\ket{0}$ and the $\ket{\pm1}$ manifold was far off resonance.
Experiments appropriate for this task could be based on static or low-frequency stress fields and would require external magnetic fields of $B_z \approx 1025$\,G to study spin-stress coupling at the ground state level anti-crossing (GSLAC), where the relevant spin sublevels are close in energy. Coupling constants $d$ and $e$ could then for example be determined via dressed state spectroscopy \cite{Barfuss2015} or the observation of stress-induced Rabi oscillations \cite{Udvarhelyi2017}.

\begin{acknowledgements}
We thank Michael Barson and Marcus Doherty for fruitful discussions. We gratefully acknowledge financial support through the NCCR QSIT, a competence center funded by the Swiss NSF and through the Swiss Nanoscience Institute (SNI). 
\end{acknowledgements}

%


\appendix
\section{Influence of NV coordinate systems on spin-oscillator coupling}
\label{app:influence_NV_coordinate_systems}
In this work, all of the employed $xyz_k$ with $k \in \{\mathrm{NV1}, \mathrm{NV2}, \mathrm{NV3}, \mathrm{NV4}\}$ are defined such that the $y$-axes lie in one of the three NV reflection planes. For example, $xyz_\mathrm{NV1}$ is determined by the unit vectors $\boldsymbol{e}_x \parallel [\bar{1}10]$, $\boldsymbol{e}_y \parallel [\bar{1}\bar{1}2]$ and $\boldsymbol{e}_z \parallel [111]$, and the corresponding spin-stress coupling Hamiltonian [see Eq.\,\eqref{eq:StressHamiltonian_NV1}] in $S=1$ basis and matrix notation reads
\begin{widetext}
	\begin{equation}
	H_\sigma = 
	\begin{pmatrix}
	M_z & \frac{1}{\sqrt 2}\left( N_x - i N_y \right) & -M_x - i M_y \\
	\frac{1}{\sqrt 2}\left( N_x + i N_y \right) & 0 & \frac{1}{\sqrt 2}\left(- N_x + i N_y \right) \\
	-M_x + i M_y & \frac{1}{\sqrt 2}\left( -N_x - i N_y \right) & M_z 
	\end{pmatrix}.
	\label{eq:StressTensor_Yaxis_MatrixNotation}
	\end{equation}
\end{widetext}
While earlier work \cite{Hughes1967, Ludlow1968} employs the same convention, it is also common to define the NV coordinate systems $xyz_k$ such that their $x$-axes are contained by a reflection plane \cite{Udvarhelyi2017, Barson2017}. A possible NV coordinate system for such a scenario would be $\boldsymbol{e}_x \parallel [11\bar{2}]$, $\boldsymbol{e}_y \parallel [\bar{1}10]$ and $\boldsymbol{e}_z \parallel [111]$, which can be obtained through a passive rotation of $xyz_\mathrm{NV1}$ by $\theta = \pi/2$ about $S_z$. The rotated interaction Hamiltonian becomes
\begin{widetext}
	\begin{equation}
	H_\sigma = 
	\begin{pmatrix}
	M_z & \frac{- i}{\sqrt 2}\left( N_x - i N_y \right) & M_x + i M_y \\
	\frac{i}{\sqrt 2}\left( N_x + i N_y \right) & 0 & \frac{-i}{\sqrt 2}\left(- N_x + i N_y \right) \\
	M_x - i M_y & \frac{-i}{\sqrt 2}\left( -N_x - i N_y \right) & M_z 
	\end{pmatrix},
	\label{eq:StressTensor_Xaxis_MatrixNotation}
	\end{equation}
\end{widetext}
and one can quickly see that rotating the NV reference frame adds phase factors of $e^{i\pi}$ and $e^{i 3 \pi/2}$ to the coupling amplitudes $M_{x,y}$ and $N_{x,y}$, respectively. 
The stress-induced spin sublevel shifts and splittings [see Eq.\,\eqref{eq:StessInducedLevelShifts_noBfield_generalExpression}] are, however, not affected. 
We therefore conclude that in the context of characterizing spin-stress coupling through the observation of stress-induced level shifts, the in-plane orientation of employed NV reference frames is of minor importance as long as they share the same quantization axis.

\section{Kelvin notation}
\label{sec:app_kelvinnotation}
For small strains or stresses, Hooke's law describes the linear stress-strain relationship
\begin{equation}
\sigma_{IJ} =\sum_{KL} C_{IJKL} \epsilon_{KL},
\label{eq:app_hookeslaw}
\end{equation}
where $C_{IJKL}$ are the components of the elastic stiffness tensor, which is a fourth rank tensor and in principle contains $3 \times 3 \times 3 \times 3 = 81$ independent elements. However, as stress and strain tensors are symmetric, this number is reduced to 36. For cubic crystals, such as diamond, symmetry arguments further reduce the number of independent elements to three \cite{Gross2012}.

To write Hooke's law in vectorial form, we employ the Kelvin notation in which \eqref{eq:app_hookeslaw} becomes 
\begin{equation}
	\boldsymbol{\tilde{\sigma}}_{XYZ} = \boldsymbol{\tilde{C}}_{XYZ}  \boldsymbol{\tilde{\epsilon}}_{XYZ}.
	\label{eq:app_hookeslawkelvin}
\end{equation}
Here, the elastic stiffness tensor reduces to a $6\times6$ tensor
\begin{equation}
\label{eq:HybridSystem_StrainStress_Quantification_StiffnessTensorVoigt}
\boldsymbol{\tilde{C}}_{XYZ} =  \begin{pmatrix} 
C_{11} & C_{12} & C_{12} & 0 & 0 & 0 \\
C_{12} & C_{11} & C_{12} & 0 & 0 & 0 \\
C_{12} & C_{12} & C_{11} & 0 & 0 & 0 \\
0 & 0 & 0 & 2 C_{44} & 0 & 0 \\
0 & 0 & 0& 0 & 2 C_{44} & 0 \\
0 & 0 & 0& 0 & 0 & 2 C_{44} \\
\end{pmatrix}, 
\end{equation}
which contains only three independent elements $\{C_{11},C_{12},C_{44}\}= \{1076,125,576\}$\,GPa for the diamond lattice symmetry \cite{Voigt1966, Gross2012, Cleland2003, Kaxiras2003}. 
Strain and stress tensors are written as the vectors
\begin{widetext}
\begin{equation}
\boldsymbol{\tilde{\epsilon}}_{XYZ} =  \left( \epsilon_{XX}, \epsilon_{YY}, \epsilon_{ZZ}, \sqrt{2}\epsilon_{YZ}, \sqrt{2}\epsilon_{XZ}, \sqrt{2}\epsilon_{XY} \right)^T
\label{eq:app_strainvectorvoigt}
\end{equation} 
and
\begin{equation}
\boldsymbol{\tilde{\sigma}}_{XYZ} =  \left( \sigma_{XX}, \sigma_{YY}, \sigma_{ZZ}, \sqrt{2}\sigma_{YZ}, \sqrt{2}\sigma_{XZ}, \sqrt{2}\sigma_{XY} \right)^T.
\label{eq:app_stressvectorvoigt}
\end{equation}
\end{widetext}

\section{Definition of rotation matrices}
\label{sec:app_definition_rotationmatrices}
The rotation matrices $\boldsymbol{R}_{\boldsymbol{n}} (\theta)$ from Tab.\,\ref{tab:CoordinateSystems} describe three-dimensional rotations by angles $\theta$ about axes indicated by the unit vectors $\boldsymbol{n} = \left(n_1, n_2, n_3\right)^T$. $\boldsymbol{R}_{\boldsymbol{n}} (\theta)$ is calculated using the relation \cite{Mehrabadi1995}
\begin{equation}
\boldsymbol{R}_{\boldsymbol{n}} (\theta) = \mathds{1} + \sin \theta \boldsymbol{N} + (1-\cos \theta) \boldsymbol{N}^2 
\label{eq:app_RotationMatrix_3x3}
\end{equation}
with 
\begin{equation}
\boldsymbol{N} = \begin{pmatrix} 0 & -n_3 & n_2 \\ n_3 & 0 & -n_1 \\ -n_2 & n_1 & 0 \end{pmatrix}. 
\label{eq:app_RotationMatrix_3x3_skewtensor}
\end{equation}
In this work, the axis of rotation $\boldsymbol{n}$ is generally defined with respect to the original, unrotated coordinate system, and $\theta$ is positive for a clockwise rotation observed along $\boldsymbol{n}$.

When working with the Kelvin notation, we employ the rotation matrices 
\begin{align}
\boldsymbol{\tilde{R}}_{\boldsymbol{n}} (\theta) & =  \tilde{\mathds{1}} + \sin \theta \boldsymbol{\tilde{N}} + (1-\cos \theta) \boldsymbol{\tilde{N}}^2 \notag \\
& + \frac{1}{3} \sin \theta (1 - \cos \theta)(\boldsymbol{\tilde{N}} + \boldsymbol{\tilde{N}}^3) \\
& + \frac{1}{6} (1-\cos \theta)^2 (\boldsymbol{\tilde{N}}^2 + \boldsymbol{\tilde{N}}^4) \notag.
\label{eq:app_RotationMatrix_6x6}
\end{align}
with $\boldsymbol{n} = (n_1, n_2, n_3)^T$ and \cite{Mehrabadi1995}
\begin{widetext}
	\begin{equation}
	\boldsymbol{\tilde{N}} = \begin{pmatrix} 0 & 0 & 0 & 0 & \sqrt 2 n_2 & -\sqrt 2 n_3 \\ 0 & 0 & 0 & -\sqrt 2 n_1 & 0 &  \sqrt 2 n_3 \\ 0 & 0 & 0 & \sqrt 2 n_1 &  -\sqrt 2 n_2 & 0 \\ 0 & \sqrt 2 n_1 & -\sqrt 2 n_1 & 0 & n_3 & -p_2 \\ -\sqrt 2 n_2 & 0 &  \sqrt 2 n_2 & -n_3 & 0 & n_1 \\ \sqrt 2 n_3 &  -\sqrt 2 n_3 & 0 & n_2 & -n_1 & 0   \end{pmatrix}.
	\label{eq:app_RotationMatrix_6x6_skewtensor}
	\end{equation}
\end{widetext}

\section{How the definition of $xyz$ influences spin-strain coupling amplitudes}

\label{sec:app_definitionxyz_couplingamplitudes}

As explained in the main text, we find the spin-strain coupling amplitudes by replacing the stress tensor components $\sigma_{IJ}$ with strain tensor components $\epsilon_{ij}$, using
\begin{equation}
\boldsymbol{\tilde{\sigma}}_{XYZ} = \boldsymbol{\tilde{C}}_{XYZ} \boldsymbol{\tilde{L}}_k^T \boldsymbol{\tilde{\epsilon}}_{xyz}^k.
\label{eq:app_StressXYZ2strainxyzNV1}
\end{equation}

Obviously, the chosen NV coordinate system $xyz_k$ has a strong impact on the resulting expressions for the spin-strain coupling amplitudes as it determines the employed rotation $\boldsymbol{\tilde{L}}_k$. To illustrate this, we consider two different NV coordinate systems that could be used as a reference frame for NV1. 
In the first case, $\boldsymbol{e}_x^{(a)} \parallel [\bar{1}10]$, $\boldsymbol{e}_y^{(a)} \parallel [\bar{1}\bar{1}2]$ and $\boldsymbol{e}_z^{(a)} \parallel [111]$ (this is $xyz_\mathrm{NV1}$ as used in the main text).
The second reference frame $xyz^{(b)}$ is obtained through rotating $xyz^{(a)}$ by an angle $\pi/2$ about the $z$-axis, which results in $\boldsymbol{e}_x^{(b)} \parallel [\bar{1}\bar{1}2]$, $\boldsymbol{e}_y^{(b)} \parallel [1\bar{1}0]$ and $\boldsymbol{e}_z^{(b)} \parallel [111]$.

We then calculate spin-strain coupling amplitudes for both NV reference frames using the rotations
\begin{align}
	\boldsymbol{\tilde{L}}_\mathrm{NV1}^{(a)} & = \boldsymbol{\tilde{R}}_{[001]} (-3\pi/4) \boldsymbol{\tilde{R}}_{[\bar{1}10]} (-\alpha_\mathrm{NV}) \notag \\
	\boldsymbol{\tilde{L}}_\mathrm{NV1}^{(b)} & = \boldsymbol{\tilde{R}}_{[001]} (-5\pi/4)\boldsymbol{\tilde{R}}_{[\bar{1}10]} (-\alpha_\mathrm{NV}). \notag
\end{align}
We find the expressions
\begin{subequations}
	\label{eq:app_straincouplingamplitudes_(a)}
	\begin{align}
	M_{x}^{(a)} & = B \left(\epsilon_{xx}^{(a)} - \epsilon_{yy}^{(a)} \right) + 2 C \epsilon_{yz}^{(a)} \\
	M_{y}^{(a)} & =-2 B \epsilon_{xy}^{(a)} - 2 C \epsilon_{xz}^{(a)} \\
	M_{z}^{(a)} & = A_1 \epsilon_{zz}^{(a)} + A_2  \left(\epsilon_{xx}^{(a)} + \epsilon_{yy}^{(a)} \right) 
	\end{align}
\end{subequations}
for reference frame $(a)$, and
\begin{subequations}
	\label{eq:app_straincoupingamplitudes_(b)}
	\begin{align}
	M_{x}^{(b)} & = - B\left(\epsilon_{xx}^{(b)} - \epsilon_{yy}^{(b)} \right) + 2 C \epsilon_{xz}^{(b)} \\
	M_{y}^{(b)} & = 2 B \epsilon_{xy}^{(b)} + 2 C \epsilon_{yz}^{(b)} \\
	M_{z}^{(b)} & = A_1 \epsilon_{zz}^{(b)} + A_2  \left( \epsilon_{xx}^{(b)} + \epsilon_{yy}^{(b)} \right) .
	\end{align}
\end{subequations}
for reference frame $(b)$. Not that both sets of coupling amplitudes depend on two strain tensors with components $\epsilon_{ij}^{(a)}$ and $\epsilon_{ij}^{(b)}$. The $N_x$ and $N_y$ terms are neglected for simplicity.

Since both reference frames share the same quantization axis, we expect identical level shifts, i.e.\,$\Delta_{\ket{\pm1}}^{(a)} = \Delta_{\ket{\pm1}}^{(b)}$. The strain-induced level shifts $\Delta_{\ket{\pm1}}^{(a)}$ and $\Delta_{\ket{\pm1}}^{(b)}$ are given by 
\begin{widetext}
	\begin{align}
		\label{eq:app_levelshifts_(a,b)}
		\Delta_{|\pm1\rangle}^{(a)}/\epsilon & = \left[ A_1 \epsilon_{zz}^{(a)} + A_2  \left(\epsilon_{xx}^{(a)} + \epsilon_{yy}^{(a)} \right) \right] \\
		& \pm \left[ B^2 (\epsilon_{xx}^{(a)} - \epsilon_{yy}^{(a)})^2 + 4 B C \epsilon_{yz}^{(a)} (\epsilon_{xx}^{(a)} - \epsilon_{yy}^{(a)}) + 4 C^2 {\epsilon_{yz}^{(a)}}^2 + 4 B^2 {\epsilon_{xy}^{(a)}}^2 + 8 B C \epsilon_{xy}^{(a)} \epsilon_{xz}^{(a)} + 4 C^2 {\epsilon_{xz}^{(a)}}^2 \right]^{1/2} \notag 
	\end{align}
	\begin{align}
		\Delta_{|\pm1\rangle}^{(b)}/\epsilon & = \left[ A_1 \epsilon_{zz}^{(b)} + a_2  \left(\epsilon_{xx}^{(b)} + \epsilon_{yy}^{(b)} \right) \right] \\
		& \pm  \left[ B^2 (\epsilon_{xx}^{(b)} - \epsilon_{yy}^{(b)})^2 - 4 B C \epsilon_{xz}^{(b)} (\epsilon_{xx}^{(b)} - \epsilon_{yy}^{(b)}) + 4 C^2 {\epsilon_{xz}^{(b)}}^2 + 4 B^2 {\epsilon_{xy}^{(b)}}^2 + 8 B C \epsilon_{xy}^{(b)} \epsilon_{yz}^{(b)} + 4 C^2 {\epsilon_{yz}^{(b)}}^2 \right]^{1/2}.  \notag
	\end{align}
\end{widetext}
and $\Delta_{\ket{\pm1}}^{(a)} \neq \Delta_{\ket{\pm1}}^{(b)}$ under the assumption $\epsilon_{ij}^{(a)} = \epsilon_{ij}^{(b)}$. 
However, the assumption $\epsilon_{ij}^{(a)} = \epsilon_{ij}^{(b)}$ is not justified, since the two strain tensors are obtained from the same stress tensor with different coordinate system transformations and thus differ as well.
To compare $\boldsymbol{\tilde{\epsilon}}_{xyz}^{(a)}$ and $\boldsymbol{\tilde{\epsilon}}_{xyz}^{(b)}$, we express them in terms of the original stress tensor $\boldsymbol{\tilde{\sigma}}_{XYZ}$ and find
\begin{widetext}
\begin{equation}
\boldsymbol{\tilde{\epsilon}}_{xyz}^{(b)} = \left( \epsilon_{yy}^{(a)}, \epsilon_{xx}^{(a)}, \epsilon_{zz}^{(a)}, -\sqrt{2} \epsilon_{xz}^{(a)}, \sqrt{2} \epsilon_{yz}^{(a)}, -\sqrt{2} \epsilon_{xy}^{(a)} \right)^T
\label{eq:app_comparisionstraintensors},
\end{equation}
\end{widetext} 
under which $\Delta_{|\pm 1\rangle}^{(a)} = \Delta_{|\pm 1\rangle}^{(b)}$ is fulfilled. 
We therefore conclude that care has to be taken when working with the spin-strain notation, since it yields reliable results for strain-induced level shifts only if the employed strain tensor in NV coordinates is derived from the stress tensor in cyrstal coordinates with the correct transformation.

\section{Engineering vs. pure strain}
\label{sec:app_engineering_vs_pure_strain}
When using Hooke's law, confusion often arises due to the difference between pure and engineering strain notations. We briefly demonstrate here that using different strain conventions does not affect the general structure of the framework presented in this paper. However, small corrections to the relations used to convert spin-stress coupling constants into their spin-strain counterparts [see Eq.\,\ref{eq:StrainCouplingConstants_LudlowForm}], as well as to the strain-induced level shifts and splittings $\Delta_{\ket{\pm1}}$ from Tab.\,\ref{tab:Summary_LevelShifts_Stress_Strain} are required.

In matrix form, the strain tensor is written as
\begin{equation}
\boldsymbol{\epsilon}_{XYZ}  = \begin{pmatrix}
\epsilon_{XX} & \epsilon_{XY} & \epsilon_{XZ} \\
\epsilon_{YX} & \epsilon_{YY} & \epsilon_{YZ} \\
\epsilon_{ZX} & \epsilon_{ZY} & \epsilon_{ZZ} \end{pmatrix}, 
\label{eq:app_straintensor_purestrain}
\end{equation}
where $\epsilon_{IJ}$ refers to the pure strain tensor components. However, to ensure the conservation of elastic energy when employing Hooke's law, the engineering strain notation is employed. A good example is the Voigt notation, in which
\begin{equation}
	\boldsymbol{\tilde{\epsilon}}_{XYZ}^V =  \left( \epsilon_{XX}, \epsilon_{YY}, \epsilon_{ZZ}, \gamma_{YZ}, \gamma_{XZ}, \gamma_{XY} \right)^T
	\label{eq:app_strainvectorvoigt_engineeringstrain}
\end{equation} 
also depends on the engineering shear strain components $\gamma_{IJ} = 2 \epsilon_{IJ}$, which are twice the pure shear strain components. However, in this work we employ the Kelvin notation, as the Voigt notation does not allow the application of standard vector operations, such as coordinate system transformations through vector rotations. In the Kelvin notation, engineering strain shows up as factors of 2 in front of the $C_{44}$ stiffness tensor components [see Eq.\,\ref{eq:HybridSystem_StrainStress_Quantification_StiffnessTensorVoigt}]. The formalism we present in this paper is therefore based on engineering strain.

This is important to note, as other works \cite{Ludlow1968, Barson2017} rely on a spin-strain coupling formalism based on pure strain, and consequently employ slightly different expressions for the conversion relations of spin-stress into spin-strain coupling constants [see Eq.\,\eqref{eq:StrainCouplingConstants_LudlowForm}] and the strain-induced level shifts $\Delta_{\ket{\pm1}}$ [see Tab.\,\ref{tab:Summary_LevelShifts_Stress_Strain}]. 
Based on the brief explanation above, we can convert our engineering strain expressions into the pure strain framework \cite{Ludlow1968, Barson2017} by replacing $C_{44} \rightarrow C_{44}/2$ in Eq.\,\eqref{eq:StrainCouplingConstants_LudlowForm}. The conversion relations for spin-stress into spin-strain coupling constants become
\begin{subequations}
	\label{eq:StrainCouplingConstants_purestrain}
	\begin{align}
	A_1 & = a_1 (C_{11} + 2C_{12}) + 2 a_2 C_{44}  \\ 
	A_2 & = a_1 (C_{11} + 2C_{12}) -  a_2 C_{44} \\ 
	B & = -b(C_{11} - C_{12}) -  c C_{44}  \\ 
	C & = \sqrt{2} b  (C_{11} - C_{12}) -  1/\sqrt{2} c C_{44} \\
	D & = -d(C_{11} - C_{12}) -  e C_{44}  \\ 
	E & = \sqrt{2} d  (C_{11} - C_{12}) - 1/\sqrt{2} e C_{44},   
	\end{align}
\end{subequations}
and yield the values
\begin{subequations}
	\begin{align}
	\label{eq:StrainCouplingSingleSpin_BendingExperiments_Staticbending_straincouplingconstants_values_purestrain}
	A_1 & = (-8.0 \pm 5.7)\text{\,GHz/strain} \\ 
	A_2 & = (-12.4 \pm 4.7)\text{\,GHz/strain}   \\ 
	B & = (-3.7 \pm 0.9)\text{\,GHz/strain}  \\ 
	C & = (11.8 \pm 1.1)\text{\,GHz/strain} 
	\end{align}
\end{subequations}
for the spin-strain coupling constants. Clearly, these values differ significantly from those given in the main text [see Eq.\eqref{eq:StrainCouplingSingleSpin_BendingExperiments_Staticbending_straincouplingconstants_values}]. Yet both strain conventions yield identical results for the strain-induced level shifts, if in the pure strain framework the $\Delta_{\ket{\pm1}}$ contain the corrected $\gamma =  2 (C_{11} + 2C_{12})/C_{44}$. It is thus of utmost importance that the employed strain framework is well defined if spin-strain coupling constants are to be compared or even determined.  

\section{Determination of spin-stress coupling constants}
\label{sec:app_determination_spinstresscouplingconstants}
\begin{figure*}[h]
	\centering
	\includegraphics[]{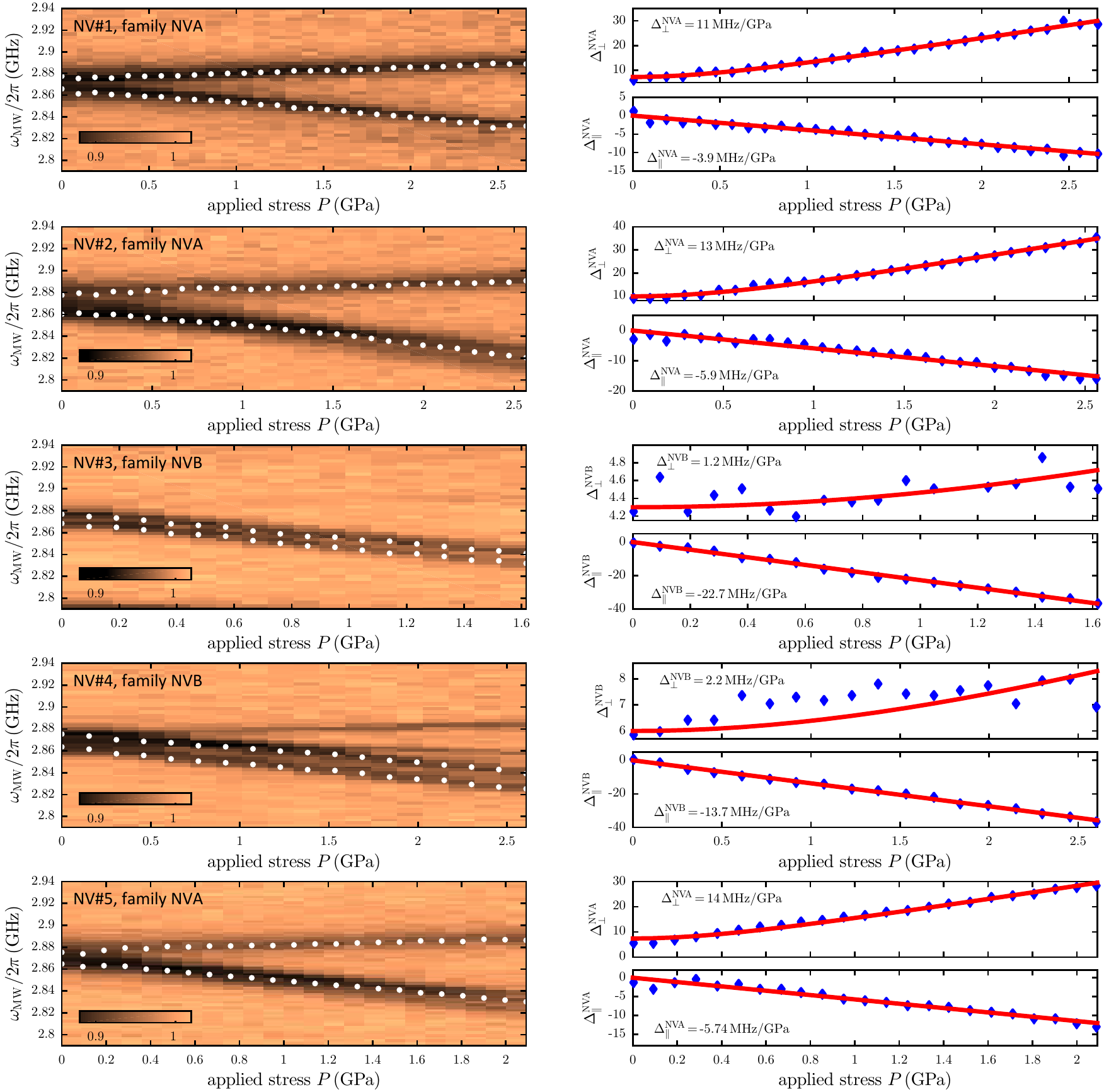}
	\caption{Reactions of different NV centers to stress induced by cantilever bending. While the three NVs from family NVA show comparable shifts and splittings, the two NVs from family NVB react quite differently to stress, causing rather large uncertainties in the stress-coupling constants. Increasing statistics, i.e.\,by studying more NV centers and their response to stress, is required. White dotted lines in the left column represent ESR peak positions, determined by fits to our original data. In the right column, blue symbols denote experimentally obtained values for level shifts and splittings and red lines are fits to these to extract spin-stress coupling constants. $\Delta_{\parallel,\perp}^{\mathrm{A},\mathrm{B}}$ are given in units of MHz. Note that our fits account for the presence of stress that was intrinsic to the sample.}
	\label{fig:extracting_spinstresscouplingconstants}
\end{figure*}

As mentioned in the main text, we analyzed the stress-induced level shifts and splittings of five different NV centers. Three of these were of orientation NVA and two belonged to NVB. Fig.\,\ref{fig:extracting_spinstresscouplingconstants} shows the taken ESR data and the extracted values of level shifts $\Delta_\parallel$ and splittings $\Delta_\perp$. White dots in the left column denote Lorentzian fits to determine the ESR dip positions. Blue dots in the right column represent calculated values for level shifts and splittings, while red lines are fits to extract level shifts and splittings per GPa of applied stress.

\newpage
\newpage





%

\end{document}